\documentclass[aps,prb,twocolumn,showpacs]{revtex4}
\usepackage[latin1]{inputenc}
\usepackage{graphicx}
\usepackage[pdftex,hypertexnames=false,pdfpagemode=UseThumbs,linkbordercolor={1 1 1}, citebordercolor={1 1 1},pdfstartview=FitH]{hyperref}
\usepackage{amsmath}
\usepackage{bbold}
\DeclareMathAlphabet{\mathpzc}{OT1}{pzc}{m}{it}
\renewcommand{\vec}[1]{\boldsymbol{#1}}
\begin{document}
\title{Electron Spin Relaxation in Graphene Nanoribbon Quantum Dots}
\author{Matthias Droth}
\author{Guido Burkard}
\affiliation{Department of Physics, University of Konstanz, 78457 Konstanz, Germany}
\begin{abstract}
Graphene is promising as a host material for electron spin qubits because of its predicted potential for long coherence times.
In armchair graphene nanoribbons (aGNRs) a small band gap is opened, allowing for electrically gated quantum dots, and furthermore the valley degeneracy is lifted. The spin lifetime $T_1$ is limited by spin relaxation, where the Zeeman energy is absorbed by lattice vibrations, mediated by spin-orbit and electron-phonon coupling. We have calculated $T_1$ by treating all couplings analytically and find that $T_1$ can be in the range of seconds for several reasons: (i) low phonon density of states away from Van Hove singularities; (ii) destructive interference between two relaxation mechanisms; (iii) Van Vleck cancellation at low magnetic fields; (iv) vanishing coupling to out-of-plane modes in lowest order due to the electronic structure of aGNRs. Owing to the vanishing nuclear spin of $^{12}$C, $T_1$ may be a good measure for overall coherence. These results and recent advances in the controlled production of graphene nanoribbons make this system interesting for spintronics applications.
\end{abstract}
\maketitle
\section{Introduction}
Graphene has attracted intense scientific interest for its mechanical, electronic, and other properties.\cite{Wallace1947,Novoselov2005,Nair2008,Meyer2007} Within the plane of its two-dimensional lattice it is extremely rigid while out-of-plane deformations are relatively soft due to the lack of a linear restoring force.\cite{Fasolino2007,Gazit2009} The absence of a band gap leads to a quasi relativistic behavior of the electrons that can be described by a Dirac-like Hamiltonian.\cite{Katsnelson2006,CastroNeto2009} However, for typical semiconductor applications such as transistors or spintronics devices, it is favorable to work with a band gap.\cite{Novoselov2004,Farmer2009,Hanson2007,Trauzettel2007} Due to Klein's paradox, a band gap is necessary to confine charge carriers electrostatically in graphene.\cite{Klein1929,Katsnelson2006} There are different situations that lead to a band gap in graphene and some of them have already been studied in view of spintronics applications.\cite{Trauzettel2007,Recher2010}

Armchair graphene nanoribbons (aGNRs) can exhibit a band gap and in addition allow for coupling of qubits in non-adjacent quantum dots (QDs).\cite{Brey2006,Han2007,Braun2011} Such a non-local coupling of qubits is ideal for fault-tolerant quantum computing and thus for scalability.\cite{Trauzettel2007,Svore2005} Over the past years, there has been substantial progress towards the goal of controlling the GNR edge termination within the production process and the controlled production of \mbox{aGNRs} might become feasible in the near future.\cite{Jiao2009,Wang2011,Zhang2013(2)}

Spintronics applications like the Loss-DiVincenzo quantum computer require spin coherence times much longer than typical operation times.\cite{Loss1998,DiVincenzo1998} When the qubit is represented by the real electron spin, carbon materials are considered promising due to the small atomic spin-orbit coupling and weak interaction with nuclear spins in carbon.\cite{Trauzettel2007,Bulaev2008,Struck2010} While the curvature significantly enhances intrinsic spin-orbit coupling and hence spin relaxation in carbon nanotubes, this effect should not occur in flat graphene.\cite{Kane2005,Min2006,Bulaev2008,Kuemmeth2008,Gmitra2009} The natural abundance of $^{13}$C, the only stable carbon isotope with a finite nuclear spin $I=1/2$ is only 1\%. The concentration of nuclear spins can be further decreased by depleting this isotope. For magnetic fields above the 10$\,$mT-regime, flip-flop processes between nuclear spins and electronic spins become suppressed due to the different magnetic moments, $\mu_{\text{B}}\gg\mu_{\text{nuc}}$. We expect that $T_2$ is dominated by $T_1$ and that the spin relaxation time is a good measure for overall coherence,  $T_1\approx T_2/2$.

\begin{figure}[b!]\centering\includegraphics[width=0.48\textwidth]{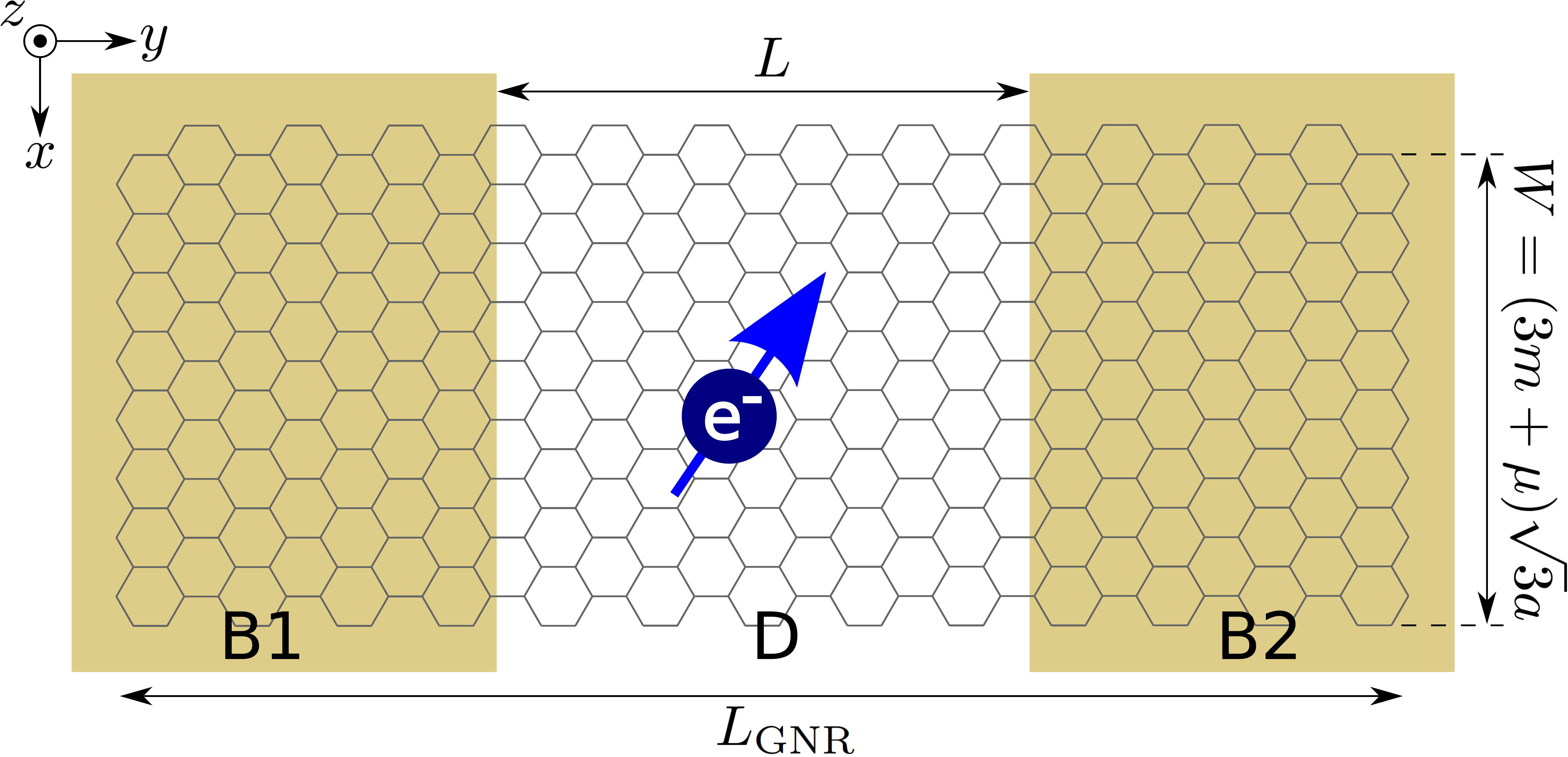}\caption{(Color online) Sketch of the system and definition of the coordinate frame. The GNR has armchair terminations in the $x$ direction. The width of the sketched aGNR is characterized by $m=3$ and $\mu=-1$, which leads to a band gap that allows for electrostatic confinement in the $y$ direction. The potential $V(y)$ defines the two barrier regions B1, B2 (shaded), and the dot region D, that lies symmetrically between the barrier regions. The interatomic distance in graphene is $a=1.42\,\text{\AA}$.}\label{pic9}
\end{figure}

In this paper, we calculate the spin relaxation time $T_1$ for electrons that are confined in an aGNR QD. The finite width of the quasi one-dimensional aGNR leads to confinement in the transverse ($x$-) direction. As we will discuss in Sec.~\ref{estates}, aGNRs of appropriate width have a band gap. This allows us to avoid Klein's paradox and confine electrons in the longitudinal ($y$-) direction by means of an electrostatic potential $V(y)$. In a perpendicular magnetic field $B\vec{e}_z$, the two possible spin states of an electron inside the QD are split by the Zeeman energy $g\mu_{\text{B}}B=\hbar\omega$, where $g=2$ is the electron $g$ factor. Figure~\ref{pic9} shows a sketch of the system.

Due to energy conservation, the Zeeman energy must be transferred to the lattice upon spin relaxation. For typical laboratory magnetic fields $B\lesssim20\,\text{T}$, the Zeeman energy corresponds to low-energy acoustic phonons at the center of the Brillouin zone.\cite{Droth2011} We consider two cases separately: (i) free and (ii) fixed boundaries. The electron-phonon coupling $\mathcal{H}_{\text{EPC}}$ comprises the deformation potential as well as the bond length change and couples in-plane vibrational modes to the electronic state. By including the spin-orbit interaction $\mathcal{H}_{\text{SOI}}$, the spin thus becomes connected to the vibrational state of the system. The coupling to the out-of-plane modes is considered, as well. Yet such a coupling either vanishes identically due to the electronic structure in aGNRs or appears only in higher order.

This paper is organized as follows. In Sec.~\ref{model}, we present our model and in Sec.~\ref{estates}, we recapitulate the bound states of aGNR QDs and explain the extended, quasi continuous states. Acoustic GNR phonons are shortly reviewed in Sec.~\ref{phon}. The effective spin-phonon coupling mechanisms that lead to $T_1^{-1}$ via Fermi's golden rule are clarified in Sec.~\ref{coupling}. In Sec.~\ref{eval}, we comment on the actual evaluation of $T_1$. The results are presented in Sec.~\ref{results} and discussed in Sec.~\ref{discuss}.
\section{Model}\label{model}
The Hamiltonian of the system is
\begin{eqnarray}
\mathcal{H}=\mathcal{H}_{\text{elec}}+\mathcal{H}_{\text{phon}}+\mathcal{H}_{\text{SOI}}+\mathcal{H}_{\text{EPC}}\,,
\end{eqnarray}
where $\mathcal{H}_{\text{elec}}$ and $\mathcal{H}_{\text{phon}}$ describe the unperturbed electronic system and the unperturbed vibrational system, respectively. The spin-orbit interaction $\mathcal{H}_{\text{SOI}}$ leads to an admixture of opposite spin states such that the electron phonon coupling $\mathcal{H}_{\text{EPC}}$ can induce a spin flip. Denoting the Fermi velocity by $v_{\text{F}}$ and the pseudospin by $\vec{\sigma}$, the unperturbed electronic part of the system obeys the Hamiltonian
\begin{eqnarray}
\mathcal{H}_{\text{elec}}\!=\!-i\hbar v_{\text{F}}
\!\begin{pmatrix}
\!\sigma_x\partial_x\!+\!\sigma_y\partial_y&0\\
0&\!-\sigma_x\partial_x\!+\!\sigma_y\partial_y
\end{pmatrix}\!+\!V(y)
\end{eqnarray}
with eigenstates $|k\rangle$. The pure vibrational modes are described by
\begin{eqnarray}
\mathcal{H}_{\text{phon}}=\sum_{\alpha,q}\hbar\omega_{\alpha,q}\left(n_{\alpha,q}+\frac{1}{2}\right)\,,
\end{eqnarray}
where the summation runs over all phonon branches $\alpha$ and wave numbers $q$. The angular frequency $\omega_{\alpha,q}$ of a vibrational mode is implicitly determined by $\alpha$ and $q$ and $n_{\alpha,q}$ is the occupation number operator. The eigenstates are the occupation number states $|n_{\alpha,q}\rangle$. 

Since $\mathcal{H}_{\text{EPC}}$ does not couple to the spin, the spin-orbit interaction $\mathcal{H}_{\text{SOI}}$ needs to be included in order to obtain a spin relaxing mechanism via admixture of electronic states.\cite{Khaetskii2001} For this admixture, we consider both bound states confined inside the dot and extended, quasi continuous states energetically above the confinement potential. 

As will be discussed in more detail, $\mathcal{H}_{\text{SOI}}$ perturbs the electron-spin product states $|k\rangle|s\rangle=|k\,s\rangle^{(0)}$, where $s=\uparrow,\downarrow$. We denote the first order perturbed states by $|k\,s\rangle$. Finally, the electron-phonon coupling leads to finite matrix elements $\langle k\!\downarrow\!\!|\mathcal{H}_{\text{EPC}}|k\!\uparrow\rangle$. This allows us to use Fermi's golden rule in order to calculate the spin relaxation rate
\begin{eqnarray}
T_1^{-1}&=&\frac{2\pi}{\hbar}\sum_{\alpha,q}|\langle k\!\downarrow,n_{\alpha,q}+1|\mathcal{H}_{\text{EPC}}|k\!\uparrow,n_{\alpha,q}\rangle|^2\nonumber\\
&&\hspace{+1.2cm}\times\rho_{\text{states}}(\hbar\omega_{\alpha,q})\,,\label{T1}
\end{eqnarray}
where $\rho_{\text{states}}(\hbar\omega_{\alpha,q})$ is the phonon density of states at the respective energy. 
The result is a function of three parameters: (i) length-to-width ratio (aspect ratio) $L/W$ of the QD, (ii) potential depth $\Delta V$ of the QD, and (iii) perpendicular magnetic field B. We find that $T_1$ can be as large as several seconds if $\rho_{\text{states}}$ is small and the two mechanisms in $\mathcal{H}_{\text{EPC}}$ interfere destructively.
\section{Electronic states}\label{estates}
\begin{figure}[t!]\centering\includegraphics[width=0.48\textwidth]{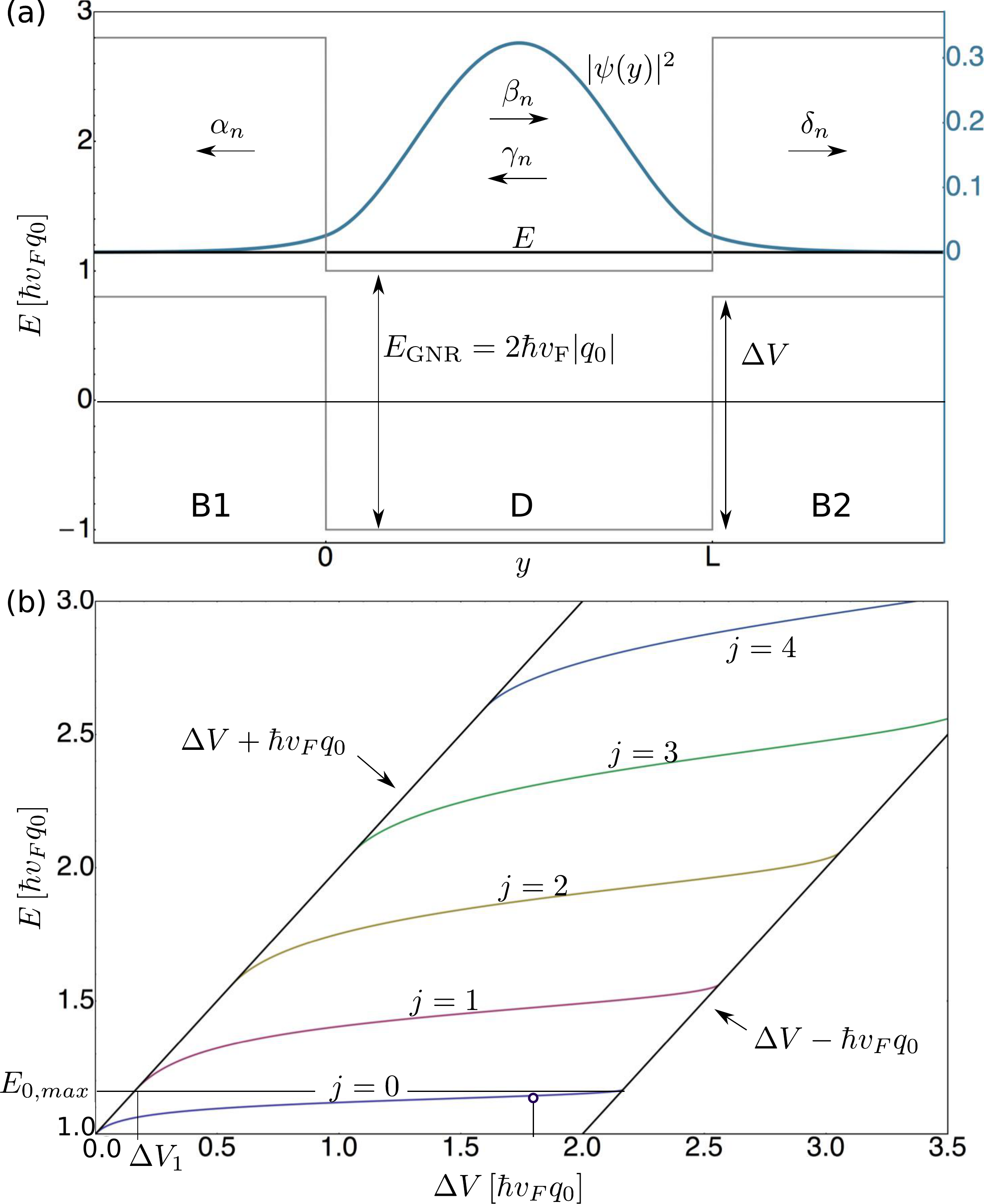}\caption{(Color online) Electron states in an aGNR QD. (a) Sketch of a bound state and (b) QD bound-state energy spectrum given by roots of Eq.~(\ref{match}). (a) Due to armchair boundaries, the minimum transverse wave number is $q_0=\pm\pi/3W$ for $\mu=\mp1$. As a consequence, the conduction band is separated from the valence band by a gap of $E_{\text{gap}}=2\hbar v_{\text{F}}|q_0|$. All energies shall be measured with respect to the middle of this band gap inside the QD region. In the barrier regions, both bands are shifted by the barrier height $\Delta V$. The resulting QD hosts at least one bound state. All bound states have the form given by Eq.~(\ref{bound}) and decay exponentially for $y\to\pm\infty$. The arrows underneath the greek letters indicate the directed character of the according part of the wave function. The plotted probability density $|\psi(y)|^2$ belongs to the lowest bound state for $L/W=5$ and $\Delta V=1.8\hbar v_{\text{F}}q_0$. (b) Bound states exist for roots of Eq.~(\ref{match}) and can be plotted in a $\Delta V$-$E$-plot. There is at least one bound state for all values of $\Delta V$. As $\Delta V$ is increased, more bound states fit into the energy gap until the lowest state can leave the QD via valence states in the barrier regions. Notably, Eq.~(\ref{match}) has exactly one root for every value of $E\geq\hbar v_{\text{F}}q_0$. We enumerate the bound states by $j=0,1,2,\ldots$. The circled position on the $j=0$ line marks the state plotted in (a). For the shown plot, the aspect ratio is $L/W=5$.}\label{pic7}
\end{figure}
Due to the aGNR edges where the wave function vanishes on both sublattices, electronic states in an aGNR have transverse wave numbers
\begin{eqnarray}
q_n=\pi(n-\mu/3)/W\,,\label{trans}
\end{eqnarray}
where $n=0,\pm1,\pm2,\ldots$ and $W=(3m+\mu)\sqrt{3}a$ is the ribbon width.\cite{Brey2006} The width depends on $m\in\mathbb{N}$ and $\mu\in\{-1,0,+1\}$. The interatomic distance is $a=1.42\,\text{\AA}$. Due to Eq.~(\ref{trans}) and $E=\pm\hbar v_{\text{F}}\sqrt{q_n^2+k^2}$, where $v_{\text{F}}$ is the Fermi velocity and $k$ is the longitudinal electronic wave number, there is a band gap $E_{\text{gap}}=2\hbar v_{\text{F}}|q_0|$. Since $E_{\text{gap}}=0$ for $\mu=0$, we assume $\mu=\pm1$ from now on. Note that $\mu$ is determined by the number of atoms across the GNR, Fig.~\ref{pic9}. Spinors with different transverse quantum number $n$ are orthogonal such that we shall focus on the lowest transverse wave number with $|q_0|=\pi/3W$. The resulting gap $E_{\text{gap}}=2\hbar v_{\text{F}}\pi/3W$ allows us to avoid Klein's paradox and confine charge carriers electrostatically in a finite square potential\cite{Trauzettel2007}
\begin{eqnarray}
V(y)=\left\{\begin{array}{lll}
0&\text{: $y\in$ D}&\text{(dot region),}\\
\Delta V&\text{: $y\in$ B1 $\cup$ B2}&\text{(barrier regions).}
\end{array}\right.
\end{eqnarray}
The barrier region B1 extends from the left end of the aGNR to $y=0$ and the barrier region B2 extends from $y=L$ to the right end. The dot region D lies symmetrically between the barrier regions. The resulting potential landscape is shown in Fig.~\ref{pic7}(a) together with a bound state, which will be discussed in the next subsection. The length of the QD is denoted by $L$ and assumed to be much smaller than the overall ribbon length, $L\ll L_{\text{GNR}}$. For concreteness, we assume an overall GNR length of $L_{\text{GNR}}=50W$.\cite{Jiao2009,Wang2011} The finite square potential needs to be considered in the electronic dispersion relation, which becomes
\begin{eqnarray}
E=V(y)\pm\hbar v_{\text{F}}\sqrt{q_0^2+k^2}\,.
\end{eqnarray}
Provided that the barrier height $\Delta V$ does not exceed a critical value $2\hbar v_{\text{F}}|q_0|+\Delta V_1$, we can easily order bound states and extended states by their energies. The critical value and $\Delta V_1$ will be explained in the next subsection - for now, we only assume that $\Delta V$ does not exceed it. Then, a state with energy $E\in[\hbar v_{\text{F}}|q_0|,\hbar v_{\text{F}}|q_0|+\Delta V]$ is \emph{bound} since its longitudinal wave number $k$ is real in the dot region and complex in the barrier regions, thus leading to an evanescent behavior. For $E>\hbar v_{\text{F}}|q_0|+\Delta V$, the longitudinal wave number is real in all regions. This leads to \emph{extended} waves. Both bound and extended states contribute to the admixture mechanism and thus shall be discussed in more detail.
\subsection{Bound states}
To describe bound states in aGNRs, one can assume an infinite ribbon.\cite{Trauzettel2007} On one hand, $L_{\text{GNR}}$ will always be finite in reality. On the other hand, bound states are mainly localized in the dot region $0\leq y\leq L$ and decay exponentially in the barrier regions, as shown in Fig.~\ref{pic7}(a). As mentioned above, we assume $L\ll L_{\text{GNR}}$, such that the overall ribbon still appears approximately infinite for bound states. This allows us to follow the description with $L_{\text{GNR}}\to\infty$ for bound states.\cite{Trauzettel2007}

Accordingly, we denote the four-component envelope wave function by
\begin{eqnarray}
\psi=(\psi_A^{(K)},\psi_B^{(K)},-\psi_A^{(K')},-\psi_B^{(K')})\label{basis}
\end{eqnarray}
and assume plane waves along the ribbon, $\psi_{n,k}^{(\pm)}(x,y)=\chi_{n,k}^{(\pm)}(x)e^{\pm iky}$, where
\begin{eqnarray}
&&\hspace{-0.5cm}\chi_{n,k}^{(+)}=a_n^{(+)}(1,z_{n,k},0,0)e^{iq_nx}+b_n^{(+)}(-z_{n,k},1,0,0)e^{-iq_nx}\nonumber\\&&\hspace{-0.5cm}+c_n^{(+)}(0,0,-z_{n,k},1)e^{iq_nx}+d_n^{(+)}(0,0,1,z_{n,k})e^{-iq_nx}
\end{eqnarray}
and
\begin{eqnarray}
&&\hspace{-0.5cm}\chi_{n,k}^{(-)}=a_n^{(-)}(z_{n,k},1,0,0)e^{iq_nx}+b_n^{(-)}(1,-z_{n,k},0,0)e^{-iq_nx}\nonumber\\&&\hspace{-0.5cm}+c_n^{(-)}(0,0,1,-z_{n,k})e^{iq_nx}+d_n^{(-)}(0,0,z_{n,k},1)e^{-iq_nx}\,.
\end{eqnarray}
With $z_{n,k}=\pm(q_n+ik)/\sqrt{q_n^2+k^2}$, and longitudinal wave numbers $k_{\text{D}}=\sqrt{(E/\hbar v_F)^2-q_n^2}$ (dot region), $\kappa_{\text{B}}=k_{\text{B}}/i=\sqrt{q_n^2-((E-e\Delta V)/\hbar v_F)^2}$ (barrier regions), bound states have the form
\begin{eqnarray}
\psi=\left\{\begin{array}{ll}
\alpha_n\chi_{n,\kappa_{\text{B}}}^{(-)}e^{\kappa_{\text{B}}y}&\text{: $y\in$ B1,}\\
\beta_n\chi_{n,k_{\text{D}}}^{(+)}e^{ik_{\text{D}}y}+\gamma_n\chi_{n,k_{\text{D}}}^{(-)}e^{-ik_{\text{D}}y}&\text{: $y\in$ D,}\\
\delta_n\chi_{n,\kappa_{\text{B}}}^{(+)}e^{-\kappa_{\text{B}}(y-L)}&\text{: $y\in$ B2.}
\end{array}\right.\label{bound}
\end{eqnarray}
The matching conditions at the interfaces B1/D and D/B2 (that is, at $y=0,L$) are discussed in Ref.~[\onlinecite{Trauzettel2007}] and can be met for roots of the transcendental equation
\begin{eqnarray}
\tan(k_{\text{D}}L)=\frac{k_{\text{D}}\kappa_{\text{B}}}{\pm\sqrt{q_n^2-\kappa_{\text{B}}^2}\sqrt{q_n^2+k_{\text{D}}^2}-q_n^2}\,.\label{match}
\end{eqnarray}
For $|q_0|=\pi/3W$ and $L/W=5$, Fig.~\ref{pic7}(b) shows these roots as a function of the barrier height $\Delta V$. There is a finite number of longitudinal excitations for any given $\Delta V$. The different bound states can be enumerated by $j=0,1,2,\ldots$ and have distinct coloring in our figure. The $j$-th bound state has $j$ nodes inside the dot region. For a given excitation, $\Delta V$ can be increased until the valence band reaches the energy of the lowest state, which can now leave the QD via valence states in the barrier regions. Note that this occurs exactly when the argument on the left-hand side of Eq.~(\ref{match}) equals a multiple of $\pi$. For the ground state, this means $k_{\text{D}}\in[0,\pi/L]$ such that the maximum ground state energy is $E_{0,\text{max}}=\hbar v_{\text{F}}\sqrt{q_0^2+(\pi/L)^2}$. States of higher energy belong to the $j$-th longitudinal excitation ($j>0$), which begins at $\Delta V_j=\hbar v_{\text{F}}(j\pi/L)$. For $\Delta V<\Delta V_1$, the ground state is the only bound state. This will be important for the evaluation of $T_1$, 
see Secs.~\ref{eval}~and~\ref{results}.

The critical value for $\Delta V$ mentioned before is $\Delta V=2\hbar v_{\text{F}}|q_0|+\Delta V_1$. If the barrier height surpasses this value, the lowest state inside the QD can leave it via valence states in the barrier region. That is the state becomes extended thus affecting the ordering of bound and extended states. Throughout this paper we assume that $\Delta V$ does not exceed this threshold such that the ground state belongs to $j=0$. 
\subsection{Extended states}
\begin{figure}[t!]\centering\includegraphics[width=0.48\textwidth]{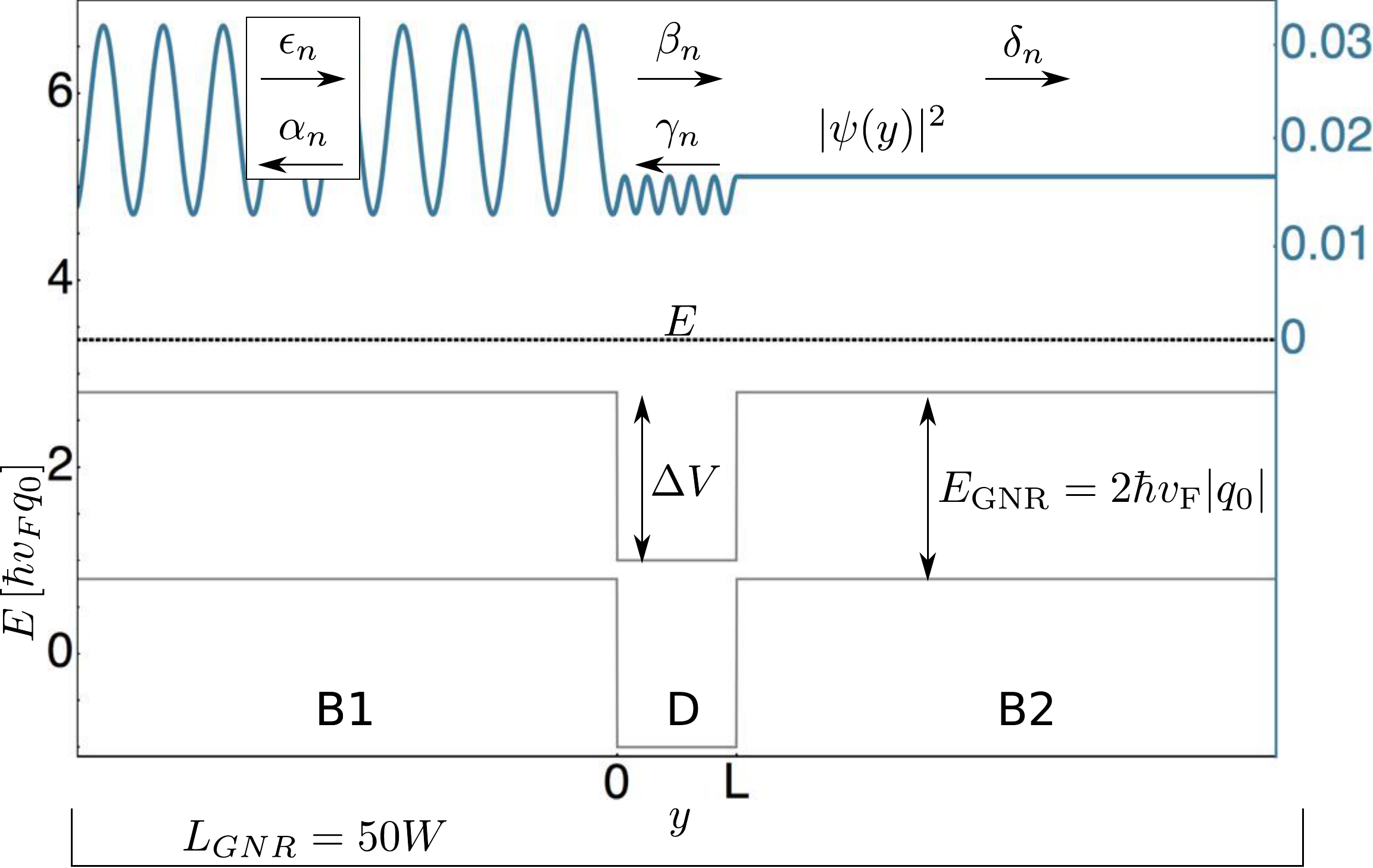}\caption{(Color online) Sketch of an extended state. The potential landscape, the aspect ratio $L/W$, and the barrier height $\Delta V$ are the same as in Fig.~\ref{pic7}(a) for bound states. The plotted probability density belongs to an extended state that is incident from the left as described by Eq.~(\ref{left}) and for which $k_{\text{EB}}=20\pi/L_{\text{GNR}}$. The arrows underneath the greek letters indicate the direction of propagation of the according part of the wave function.}\label{pic8}
\end{figure}
We assume $L_{\text{GNR}}=50W$ for the overall length of the GNR such that possible wave numbers are $k_{\text{EB}}=0,\pm2\pi/L_{\text{GNR}},\ldots,\pm\pi/a$ with lattice constant $a$. Since energy is conserved, the wave number becomes
\begin{eqnarray}
k_{\text{ED}}=\sqrt{\left(\sqrt{q_n^2+k_{\text{EB}}^2}+\Delta V/\hbar v_F\right)^2-q_n^2}
\end{eqnarray}
in the dot region. Depending on the sign of $k_{\text{EB}}$, the state is incident from $y<0$, leading to
\begin{eqnarray}
\psi\!=\!\left\{\begin{array}{ll}
\epsilon_n\chi_{n,k_{\text{EB}}}^{(+)}e^{ik_{\text{EB}}y}+\alpha_n\chi_{n,k_{\text{EB}}}^{(-)}e^{-ik_{\text{EB}}y}&\text{: $y\in$ B1,}\\
\beta_n\chi_{n,k_{\text{ED}}}^{(+)}e^{ik_{\text{ED}}y}+\gamma_n\chi_{n,k_{\text{ED}}}^{(-)}e^{-ik_{\text{ED}}y}&\text{: $y\in$ D,}\\
\delta_n\chi_{n,k_{\text{EB}}}^{(+)}e^{ik_{\text{EB}}(y-L)}&\text{: $y\in$ B2,}
\end{array}\right.\label{left}
\end{eqnarray}
see Fig.~\ref{pic8}, or it is incident from  $y>L$, which is described by
\begin{eqnarray}
\psi\!=\!\left\{\begin{array}{ll}
\!\alpha_n\chi_{n,k_{\text{EB}}}^{(-)}e^{-ik_{\text{EB}}y}&\!\!\!\text{: $y\!\in$B1,}\\
\!\beta_n\chi_{n,k_{\text{ED}}}^{(+)}e^{ik_{\text{ED}}y}+\gamma_n\chi_{n,k_{\text{ED}}}^{(-)}e^{-ik_{\text{ED}}y}&\!\!\!\text{: $y\!\in$D,}\\
\!\delta_n\chi_{n,k_{\text{EB}}}^{(+)}e^{ik_{\text{EB}}(y-L)}+\epsilon_n\chi_{n,k_{\text{EB}}}^{(-)}e^{-ik_{\text{EB}}(y-L)}&\!\!\!\text{: $y\!\in$B2.}
\end{array}\right.\hspace{-0.61cm}\nonumber\\
\hfill\label{right}
\end{eqnarray}
The matching conditions at $y=0,L$ can always be met. In contrast to bound states, extended states are propagating waves in the barrier regions.
\section{Acoustic GNR phonons}\label{phon}
\begin{figure}[t!]\centering\includegraphics[width=0.48\textwidth]{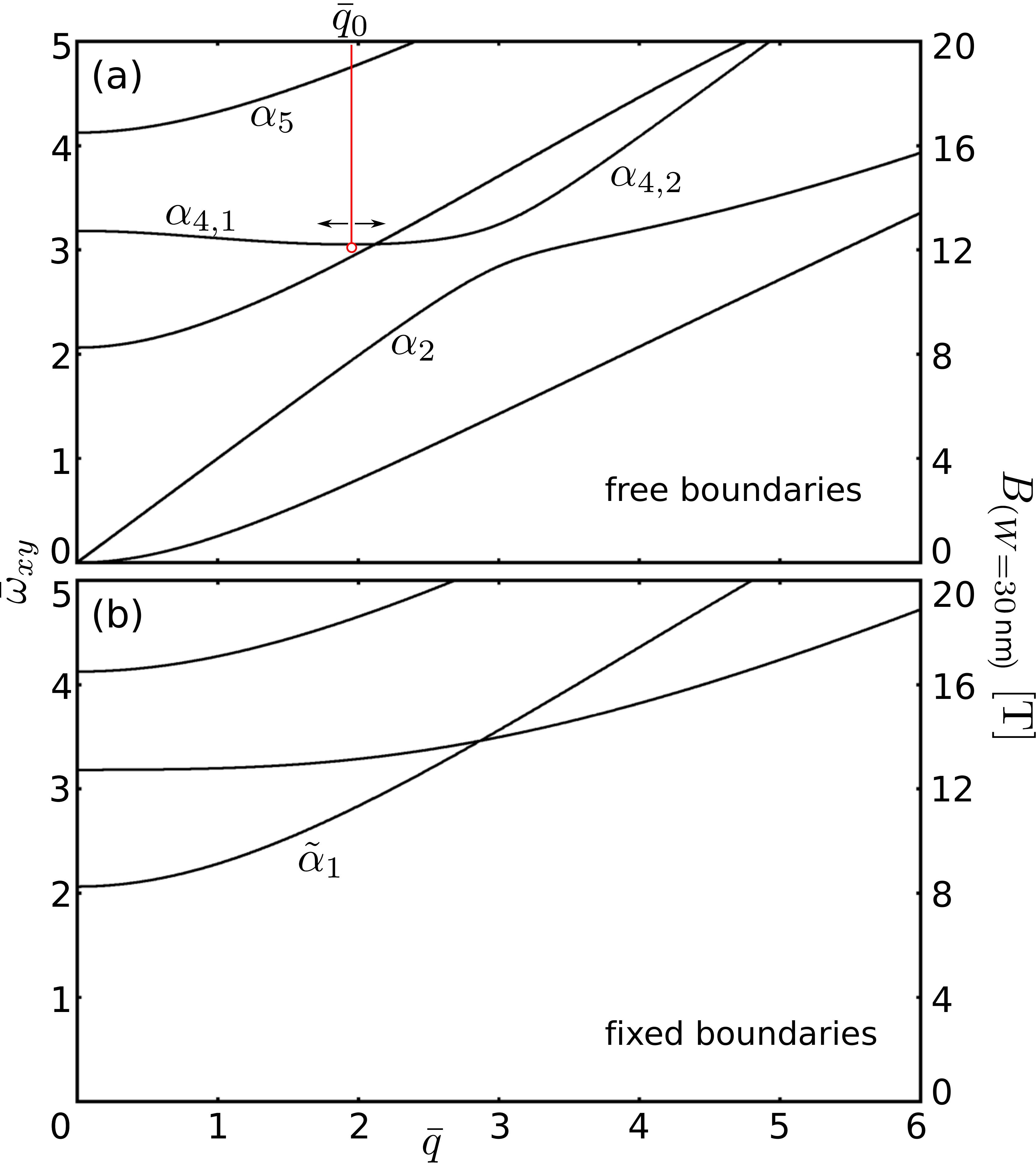}\caption{(Color online) Phonon dispersion for (a) free and (b) fixed boundaries. The dimensionless frequency $\bar{\omega}_{xy}$ is connected to the physical frequency by $\bar{\omega}_{xy}=\omega\sqrt{\rho/\mathpzc{E}}\,W$, where the radicand contains elastic constants listed in Table~\ref{constants}. We restrict our interest to the frequency range $\bar{\omega}_{xy}\in[0,5]$ since for $W=30\,\text{nm}$, the upper bound already relates to a magnetic field of $20\,\text{T}$. The scale on the right-hand side shows the magnetic field for $W=30\,\text{nm}$. Due to parity with respect to $x$, only the labeled branches ($\alpha_i$ for free and $\tilde{\alpha}_1$ for fixed boundaries) assist in spin relaxation.  (a) The phonon spectrum is gapless for free boundaries. The branch $\alpha_4$ has a minimum and hence a diverging density of states for finite $\bar{q}$. Its constituent parts $\alpha_{4,1}$ and $\alpha_{4,2}$ shall be treated separately. (b) Fixed edges lead to gapped phonon spectrum. For $W=30\,\text{nm}$, this gap corresponds to $8.25\,\text{T}$. In our range of interest, the branch labeled $\tilde{\alpha}_1$ provides the only channel for spin relaxation.}\label{pic6}
\end{figure}
The phonon energies we are interested in need to match the Zeeman splitting, $\hbar\omega=g\mu_{\text{B}}B$, where $\omega$ is the phonon frequency, $g$ the electron g factor, and $\mu_{\text{B}}$ denotes Bohr's magneton. For typical laboratory magnetic fields ${B\lesssim20\,\text{T}}$, this implies low-energy acoustic phonons at the center of the Brillouin zone, which can be modeled by continuum mechanics.\cite{Droth2011,LandauLifshitz} In this model, deformations are described by the displacement field $\vec{u}(\vec{r})$. While the components $u_{xz}$ and $u_{yz}$ of the strain tensor $u_{ik}=(\partial_iu_k+\partial_ku_i)/2$ are known to vanish for thin plates in general, the monatomic thickness  of graphene implies that $u_{zz}$ must vanish, as well. With $u_{iz}\equiv0$, the elastic Lagrangian density of monolayer graphene is given by\cite{Suzuura2002,Mariani2009,Droth2011}
\begin{eqnarray}
\mathcal{L}=\mathcal{T}-\mathcal{V}=\frac{\rho}{2}\dot{\vec{u}}^2-\frac{\kappa}{2}(\triangle u_z)^2-\frac{B+\mu}{2}u_{ii}^2+\mu u_{ik}^2\,,\label{lagrangian}
\end{eqnarray}
where $\triangle=\partial_x^2+\partial_y^2$, the sum convention with $u_{ii}=u_{xx}+u_{yy}+u_{zz}$ and $u_{ik}^2=u_{xx}^2+u_{xy}^2+\cdots$ has been used, $\rho$ is the mass density, and $\kappa$ is the bending rigidity. The bulk ($B$) and shear ($\mu$) moduli can be expressed by Poisson's ratio $\sigma$ and Young's modulus $\mathpzc{E}$. The numerical values of elastic and other constants we use are listed in Table~\ref{constants}. Equation~(\ref{lagrangian}) shows that in-plane vibrations $\vec{u}_{\parallel}$ decouple from out-of-plane vibrations $\vec{u}_{\bot}$. By assuming  $u_i(x,y)=f^i(x)\exp[i(qy-\omega t)]$ for a single mode and imposing free or fixed conditions as discussed in Ref.~[\onlinecite{Droth2011}], we obtain the according phonon dispersions (Fig.~\ref{pic6}) and the explicit displacement fields. The latter can be quantized and then take the form
\begin{eqnarray}
&&\vec{u}_{||}=\sum_{\alpha,q}r_{\alpha,q}(f_{\alpha,q}^x\vec{e}_x+f_{\alpha,q}^y\vec{e}_y)e^{iqy}\,,\\
&&\vec{u}_{\bot}=\sum_{\alpha,q}r_{\alpha,q}f_{\alpha,q}^z\vec{e}_ze^{iqy}\,,
\end{eqnarray}
where $q$ is the phonon wave number, $\alpha$ labels the phonon branch, and
\begin{eqnarray}
r_{\alpha,q}=\sqrt{\hbar/(2\rho LW\omega_{\alpha,q})}(b_{\alpha,q}+b_{\alpha,-q}^{\dagger})\label{normal}
\end{eqnarray}
is the normal coordinate. The operator $b_{\alpha,q}$ ($b_{\alpha,q}^{\dagger}$) annihilates (creates) a phonon on branch $\alpha$ with wave number~$q$.

As discussed in the following section, we can neglect coupling to out-of-plane modes and focus on in-plane modes. The dimensionless frequency of in-plane modes $\bar{\omega}_{xy}$ is related to the physical frequency by $\bar{\omega}_{xy}=\omega\sqrt{\rho/\mathpzc{E}}\,W$. In the continuum model, all branches extend to infinity but we are only interested in the range $\bar{\omega}_{xy}\in[0,5]$ since for a typical GNR width of $W=30\,\text{nm}$, $\bar{\omega}_{xy}=5$ relates to a magnetic field of $20\,\text{T}$. The dimensionless wave number $\bar{q}$ is obtained from the physical wave number via $\bar{q}=qW$. For symmetry reasons explained in Sec.~\ref{eval}, not all branches contribute to spin relaxation but only those which are explicitly labeled in Fig.~\ref{pic6}.

In the case of free boundaries, the branches $\alpha_2$, $\alpha_{4}$, and $\alpha_5$ are relevant. While $\alpha_2$ extends throughout the considered interval, $\alpha_5$ only exists above $\bar{\omega}_{xy}=4.16$, and $\alpha_4$ needs further discussion. Its minimum is $3.05$ and occurs at a finite value $\bar{q}_0$ where the density of states has a Van Hove singularity. We consider two parts: for $\bar{q}<\bar{q}_0$, we label the branch $\alpha_{4,1}$ and its label for $\bar{q}>\bar{q}_0$ is $\alpha_{4,2}$. The range of $\alpha_{4,1}$ is $\bar{\omega}_{xy}\in[3.05,3.18]$ and $\alpha_{4,2}$ extends from its minimum to infinity.

For fixed boundaries, we only need to consider the branch $\tilde{\alpha}_1$. It extends from $\bar{\omega}_{xy}=2.06$ to infinity. We emphasize that for a typical GNR width of $30\,\text{nm}$, single-phonon processes do not occur up to $8.25\,\text{T}$ as there are no phonons below $\bar{\omega}_{xy}=2.06$ for fixed boundaries.
\section{Coupling mechanisms}\label{coupling}
Phonons do not couple to the electron spin directly. The relevant mechanism usually involves the spin-orbit interaction.\cite{Khaetskii2001,Bulaev2008,Struck2010} In graphene, the spin-orbit interaction is given by
\begin{eqnarray}
\mathcal{H}_{\text{SOI}}&=&\mathcal{H}_I+\mathcal{H}_R\nonumber\\
&=&\lambda_I\tau_z\sigma_zs_z+\lambda_R(\tau_z\sigma_xs_y-\sigma_ys_x)
\end{eqnarray}
and we will consider it in order to obtain an indirect spin-phonon coupling.\cite{Kane2005,Min2006,Gmitra2009} The intrinsic (Dresselhaus) term $\mathcal{H}_I$ has coupling strength $\lambda_I$ and the Rashba (or extrinsic) term $\mathcal{H}_R$ couples with strength $\lambda_R$. The valley is denoted by $\tau_z$, pseudospin by $\vec{\sigma}$, and real spin by $\vec{s}$. In the following, we show how the real electron spin can be connected to the vibrational state of the system by taking the spin-orbit interaction into account.
\subsection{Coupling to in-plane modes}
In first order perturbation theory, $\mathcal{H}_R$ corrects the electron-spin product states $|k\rangle|\!\uparrow\rangle=|k\!\uparrow\rangle^{(0)}$ to
\begin{eqnarray}
|k\!\uparrow\rangle=|k\!\uparrow\rangle^{(0)}+\sum_{k'\neq k}|k'\!\!\downarrow\rangle^{(0)}\frac{^{(0)}\langle k'\!\!\downarrow\!|\mathcal{H}_R|k\!\uparrow\rangle^{(0)}}{E_k-E_k'+g\mu_{\text{B}}B}\,,\label{perturb}
\end{eqnarray}
and $|k\!\downarrow\rangle$ accordingly. We emphasize that the summation index $k'$ runs over both bound and extended states. The potential depth and the aspect ratio determine how many bound states exist, Fig.~\ref{pic7}(b). For extended states, we consider all wave numbers inside the first Brillouin zone, $k_{\text{EB}}=0,\pm2\pi/L_{\text{GNR}},\ldots,\pm\pi/a$.

The second term in Eq.~(\ref{perturb}) admixes states with opposite spin such that the electron-phonon coupling $\mathcal{H}_{\text{EPC}}$ can induce a spin flip\cite{Khaetskii2001}
\begin{eqnarray}
&&\hspace{-1.1cm}\langle k\!\downarrow|\mathcal{H}_{\text{EPC}}|k\!\uparrow\rangle\nonumber\\
&&\hspace{-0.9cm}=\!\sum_{k'\neq k}\!\left[\frac{(\mathcal{H}_{\text{EPC}})_{kk'}(\mathcal{H}_R)_{k'k}^{\downarrow\uparrow}}{E_k-E_{k'}+g\mu_{\text{B}}B}\!+\!\frac{(\mathcal{H}_{\text{EPC}})_{k'k}(\mathcal{H}_R)_{kk'}^{\downarrow\uparrow}}{E_k-E_{k'}-g\mu_{\text{B}}B}\right]\!,\label{admix}
\end{eqnarray}
where we denote the numerator in Eq.~(\ref{perturb}) as $(\mathcal{H}_R)_{k'k}^{\downarrow\uparrow}$ and the spin-conserving transitions of $\mathcal{H}_{\text{EPC}}$ accordingly. We find that for a given $k'$, the two terms in Eq.~(\ref{admix}) exactly cancel each other at $B=0$. This effect is known as Van Vleck cancellation and is expected for time-reversal-symmetric systems.\cite{VanVleck1940} Moreover, $(\mathcal{H}_R)_{kk'}^{\downarrow\uparrow}$ vanishes if both $k$ and $k'$ represent bound states and the longitudinal excitation indices $j_k$, $j_{k'}$ (see Fig.~\ref{pic7}) are both even or both odd. In the electron phonon coupling Hamiltonian $\mathcal{H}_{\text{EPC}}$, we consider the deformation potential $\mathcal{H}_{\text{DP}}$ as well as bond length change $\mathcal{H}_{\text{BLC}}$:
\begin{eqnarray}
&&\mathcal{H}_{\text{EPC}}=\mathcal{H}_{\text{DP}}+\mathcal{H}_{\text{BLC}}\,,\label{epc}\\
&&\mathcal{H}_{\text{DP}}=g_1\vec{\nabla}\cdot\vec{u}_{\parallel}\,,\nonumber\\
&&\mathcal{H}_{\text{BLC}}=g_2\begin{pmatrix}
0&\!A_x\!-\!iA_y\!&0&0\\
\!A_x\!+\!iA_y\!&0&0&0\\
0&0&0&\!A_x\!+\!iA_y\!\\
0&0&\!A_x\!-\!iA_y\!&0\\
\end{pmatrix}\,,\nonumber
\end{eqnarray}
where $g_{1,2}$ are coupling constants, $(A_x,A_y)=(u_{xx}-u_{yy},-2u_{xy})$, and the basis of Eq.~(\ref{basis}) has been used.\cite{Suzuura2002,Mariani2009,CastroNeto2009}
\subsection{Vanishing out-of-plane deflection coupling}
Low-energy acoustic phonons at the center of the Brillouin zone have a wavelength much larger than the lattice constant and produce a local tilt of the GNR. In the local ribbon frame $\Sigma'$ where $\vec{n}=\vec{e}_z'$ is the vector normal to the ribbon plane the local spin matrix is described by $s_z'=s_z-\partial_xu_zs_x-\partial_yu_zs_y$. As a consequence, the intrinsic spin-orbit interaction
\begin{eqnarray}
\mathcal{H}_I=\lambda_I\tau_z\sigma_z(s_z-\partial_xu_zs_x-\partial_yu_zs_y)\label{deflect}
\end{eqnarray}
becomes dependent on out-of-plane phonons such that these could flip the spin. This is known as deflection coupling.\cite{Rudner2010,Struck2010} However, there is a proportionality to $\tau_z$ in Eq.~(\ref{deflect}). Since the electronic states we use have the property that the wave function has equal weight on each valley,\cite{comment} the contributions from $K$ and $K'$ add up to zero and the deflection coupling between spin and out-of-plane modes vanishes.

If the spin-orbit admixed states of Eq.~(\ref{perturb}) are used, there is a finite overlap only between both admixed parts such that the resulting mechanism is proportional to $\lambda_I\lambda_R^2$ and hence negligible.

Compared to in-plane phonons, both deformation potential and bond length change appear only in higher order such that we neglect these mechanisms for out-of-plane phonons.
\section{Evaluation of $T_1$}\label{eval}
Using Eq.~(\ref{T1}), we calculate the spin relaxation rate for the electron in the lowest bound state (ground state) of the QD. For concreteness, we assume $\mu=-1$ in Eq.~(\ref{trans}). According to Eqs.~(\ref{bound}), (\ref{left}), and (\ref{right}), both bound and extended states have a finite probability density in the barrier regions. However, bound states are localized in the dot region and decay exponentially in the barrier regions. In particular the ground state, plotted in Fig.~\ref{pic7}(a), has a very low probability density in the barrier regions. Its overlap with another bound state in the barrier regions is negligible. Only high-energy extended states have a significant contribution outside the dot region. Yet the overlap of an extended state with the ground state outside the dot region will still be small and since they are energetically far apart, the contribution from the barrier regions can be neglected. As a consequence, we can restrict the integrals in Eq.~(\ref{admix}) to the dot region.

As discussed in Sec.~\ref{phon}, we consider phonons with free boundaries as well as phonons with fixed boundaries. Not all branches contribute to spin relaxation: Because of mirror symmetry with respect to $x=0$, $\mathcal{H}_{\text{BLC}}$ and $\mathcal{H}_{\text{DP}}$ are even or odd in $x$, depending on what phonon branch they belong to. Due to their similar form,\cite{comment2} the mechanisms are either both even or both odd for a given branch. The $x$ dependencies of the electronic states in the matrix element $(\mathcal{H}_{\text{EPC}})_{k'k}$ cancel out, $e^{i(q_0-q_0)x}=1$, such that the $x$ integral vanishes if Eq.~(\ref{epc}) is odd in $x$. The branches $\alpha_2$, $\alpha_4$, and $\alpha_5$ in Fig.~\ref{pic6}(a) and $\tilde{\alpha}_1$ in Fig.~\ref{pic6}(b) have couplings $\mathcal{H}_{\text{EPC}}$ that are even in $x$ and hence can relax the spin.

For a given relaxation channel $(\alpha,q)$, both mechanisms $\mathcal{H}_{\text{DP}}$ and $\mathcal{H}_{\text{BLC}}$ are combined in Eq.~(\ref{T1}) coherently. Moreover, the couplings via bound states and extended states in Eq.~(\ref{perturb}) are added up in a coherent way. We are interested in the relaxation of the spin in the ground state, which corresponds to $j=0$ in Fig.~\ref{pic7}(b) and hence restrict the barrier height to $\Delta V\in[0,\,2\hbar v_{\text{F}}q_0+\Delta V_1]$. If $\Delta V$ exceeds this upper bound, valence states become available in the barrier regions and the lowest state inside the QD can leave the dot region. For $\Delta V<\Delta V_1$ on the other hand, the ground state is the only bound state such that the perturbation in Eq.~(\ref{perturb}) comes about only due to extended states, which fully determine the spin relaxation in this case.

For spin relaxation, Eq.~(\ref{T1}) is proportional to $n_{\alpha,q}+1$ and we assume $n_{\alpha,q}=0$, i.\ e., sufficiently low temperature, $k_{\text{B}}T\ll\hbar\omega=g\mu_{\text{B}}B$. By $k_{\text{B}}$ we denote Boltzmann's constant. Assuming a magnetic field of $B=1\,\text{T}$, this means $T\ll1.3\,\text{K}$. For $T\gtrsim15\,\text{K}$, spontaneous emission can be neglected since $n_{\alpha,q}\gg1$ and one obtains the spin relaxation by multiplying our results with the expectation value of the Bose distribution
\begin{eqnarray}
\langle n_{\alpha,q}(B,T)\rangle=\left(e^{\frac{g\mu_{\text{B}}B}{k_{\text{B}}T}}-1\right)^{-1}\,.
\end{eqnarray}

The spin relaxation time $T_1$ is a good measure for overall coherence when pure dephasing, which comes from coupling to nuclear spins, is negligible. Due to the low density of nuclear spins in natural carbon and the very different magnetic moments $\mu_{\text{B}}\gg\mu_{\text{nuc}}$, we expect that flip-flop processes with nuclear spins can be neglected for magnetic fields above 10$\,$mT. For a typical GNR width of $W=30\,\text{nm}$, 10$\,$mT correspond to $\bar{\omega}_{xy}=0.0025$. As a consequence, we restrict our calculations to the interval $\bar{\omega}_{xy}\in[0.0025,5]$. The upper bound corresponds to a magnetic field of 20$\,$T. All plots that show rates are cut off at these boundaries.
\section{Results}\label{results}
To calculate $T_1$, we need to use the specific values of the elastic constants that define the phonon spectrum. Young's modulus for the two-dimensional lattice of graphene is obtained by multiplying the bulk value with the thickness associated with graphene, $\mathpzc{E}=\mathpzc{E}_{\text{3D}}h$, where $h=3.4\,\text{\AA}$. For further discussion of the elastic constants, we refer to Ref.~[\onlinecite{Droth2011}]. Table~\ref{constants} gives an overview of the constants we use in our calculation. The Rashba-type spin-orbit coupling is linear in the electric field and thus can be adjusted by an external electric field or by using a suitable substrate.\cite{Min2006}
\begin{table}[h!]
\centering
\begin{tabular}{llll}
\hline
$\sigma=0.16$&{\fontsize{6}{12}\selectfont[\onlinecite{Lee2008,Faccio2009,Kudin2001}]}$\hspace{+0.1cm}$&$g_1=30\,\text{eV}$&{\fontsize{6}{12}\selectfont[\onlinecite{Suzuura2002,Mariani2009,Struck2010}]}\\
$\mathpzc{E}=3.4\,\text{TPa$\,$\AA}$&{\fontsize{6}{12}\selectfont[\onlinecite{Lee2008,Faccio2009,Kudin2001}]}&$g_2=1.5\,\text{eV}$&{\fontsize{6}{12}\selectfont[\onlinecite{Mariani2009,Struck2010}]}\\
$B=12.6\,\text{eV/\AA}^2$&{\fontsize{6}{12}\selectfont[\onlinecite{Kudin2001,Gazit2009}]}&$\lambda_{\text{R}}=40\times10^{-6}\,\text{eV}$&{\fontsize{6}{12}\selectfont[\onlinecite{Min2006,Gmitra2009,Struck2010}]}\\
$\mu=9.1\,\text{eV/\AA}^2$&{\fontsize{6}{12}\selectfont[\onlinecite{Kudin2001,Gazit2009}]}&$v_F=8.8\times10^5\,\text{m/s}$&{\fontsize{6}{12}\selectfont[\onlinecite{CastroNeto2009,Gmitra2009,Struck2010}]}\\
$\rho=7.61\times10^{-7}\,\text{kg/m}^2$&{\fontsize{6}{12}\selectfont[\onlinecite{comment3}]}\\
\hline
\end{tabular}
\caption{Numerical values of the parameters we use in our calculation.} \label{constants}
\end{table}

The spin relaxation time $T_1$ depends on three parameters: (i) the aspect ratio $L/W$ of the QD, (ii) the potential depth $\Delta V$ of the QD, and (iii) the applied perpendicular magnetic field $B\propto\bar{\omega}_{xy}$. Moreover, the phonon spectrum and hence the spin relaxation depends on the mechanical boundary conditions. We discuss free boundary conditions separately from fixed boundaries.
\subsection{Free boundary conditions}
\begin{figure}[t!]\centering\includegraphics[width = 0.48\textwidth]{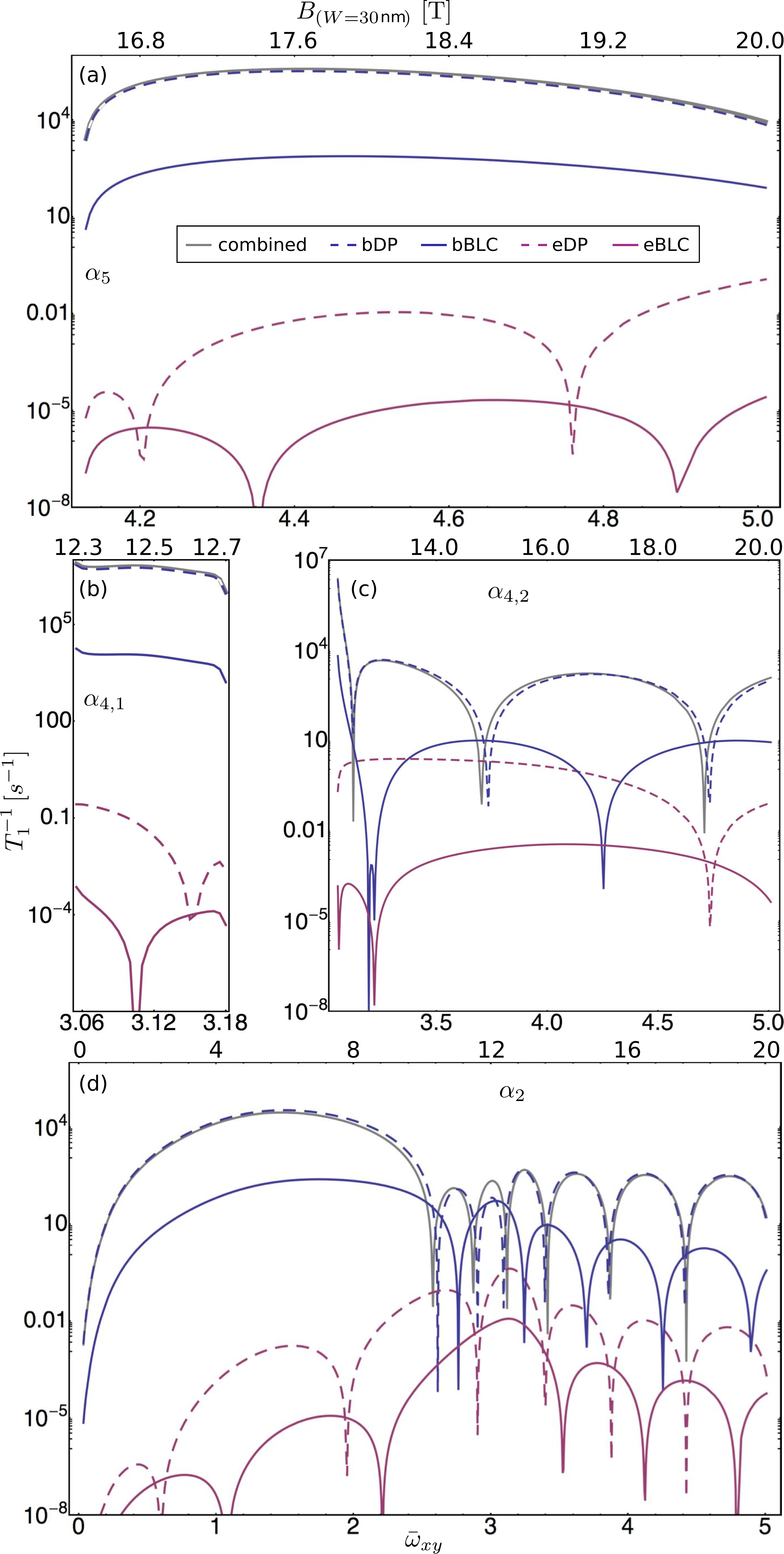}\caption{(Color online) Partial rates for various relaxation channels. For $L/W=5$ and $\Delta V=1.8\hbar v_{\text{F}}q_0$,  all contributions to the four relaxation channels $\alpha_5$ (a), $\alpha_{4,1}$ (b), $\alpha_{4,2}$ (c), and $\alpha_2$ (d) are shown. The contributions stem from $\mathcal{H}_{\text{DP}}$ with admixture of bound (labeled ``bDP") states or extended states (``eDP") and  from $\mathcal{H}_{\text{BLC}}$ with the same admixtures (``bBLC" and ``eBLC", respectively). These contributions are added up coherently to the ``combined" relaxation of the respective channel.}\label{pic1}
\end{figure}
For symmetry reasons explained above, only the phonon branches with labels $\alpha_2$, $\alpha_4$ (consisting of parts $\alpha_{4,1}$ and $\alpha_{4,2}$), and $\alpha_5$ in Fig.~\ref{pic6}(a) need to be considered. The respective rates of these relaxation channels are shown in Fig.~\ref{pic1} for an aspect ratio of $L/W=5$ and a barrier height of $\Delta V=1.8\hbar v_{\text{F}}q_0$. The allocation of panels to branches is as follows: Fig.~\ref{pic1}(a) belongs to branch $\alpha_5$, \ref{pic1}(b) to $\alpha_{4,1}$, \ref{pic1}(c) to $\alpha_{4,2}$, and \ref{pic1}(d) to $\alpha_2$. Each panel shows four separate contributions to $T_1^{-1}$ that come about from the two mechanisms in Eq.~(\ref{epc}) and the admixture of bound states or of extended states in Eq.~(\ref{perturb}) for each mechanism. The coherent sum of all four contributions is displayed by the gray line. The deformation potential usually dominates over the bond length change since its coupling constant is 20 times larger, Table~\ref{constants}. For $\Delta V=1.8\hbar v_{\text{F}}q_0$, extended states are energetically far away from the ground state such that the contribution from the deformation potential with admixture of bound states dominates in Fig.~\ref{pic1}. Oscillations in individual rates may be due to the phonon phase $e^{iqy}$ that is integrated with the matrix elements $(\mathcal{H}_{\text{EPC}})_{k'k}$ and rotates according to the phonon dispersion when $\omega$ is changed. Figures \ref{pic1}(c) and \ref{pic1}(d) show that the matrix elements $(\mathcal{H}_{\text{DP}})_{k'k}$ and $(\mathcal{H}_{\text{BLC}})_{k'k}$ may interfere destructively, thus decreasing $T_1^{-1}$ by several orders of magnitude, yet typically not to zero.

In all these plots, the bottom scale shows $\bar{\omega}_{xy}$ and the top scale shows the magnetic field $B$ that corresponds to $\bar{\omega}_{xy}$, assuming a width of $W=30\,\text{nm}$. Note, that $T_1^{-1}$ does not depend on $B$ and $W$ separately, but only on the product $BW\propto\omega W\propto\bar{\omega}_{xy}$.

\begin{figure}[t!]\centering\includegraphics[width = 0.48\textwidth]{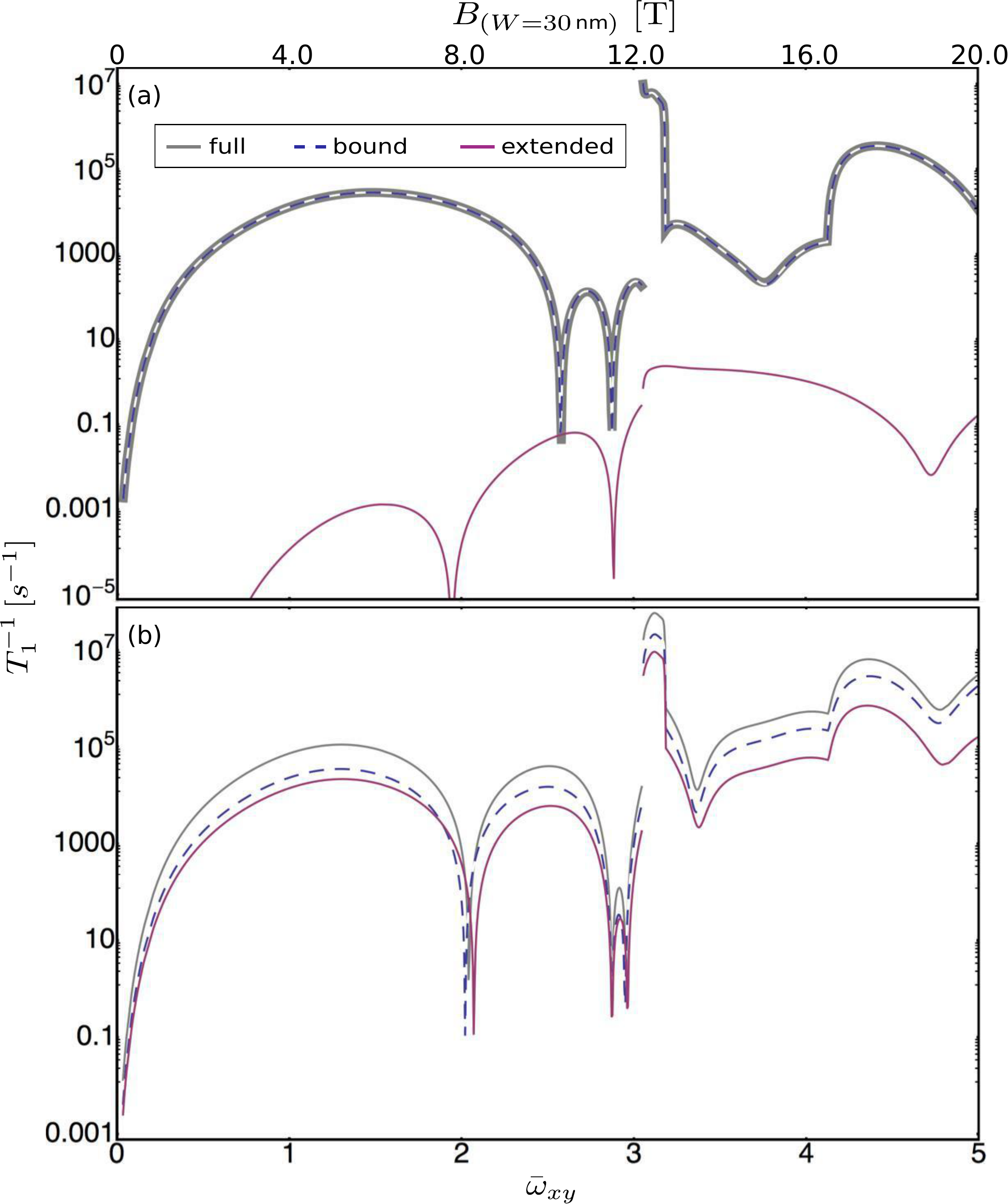}\caption{(Color online) The relaxation rates for different dot depths $\Delta V$. By summing up the combined relaxation rates (see Fig.~\ref{pic1}) of all channels available for a certain $\bar{\omega}_{xy}$, the full relaxation rate (gray line) is obtained. The lines labeled ``bound" and ``extended" are obtained in a similar way by considering only bound or extended states, respectively. At $\bar{\omega}_{xy}=3.05$, $T_1^{-1}$ is discontinuous due to the advent of the relaxation channel $\alpha_4$ that has a diverging density of states at this point, Fig.~\ref{pic6}(a). (a) accords to parameters $L/W=5$ and $\Delta V=1.8\hbar v_{\text{F}}q_0$ as in Fig.~\ref{pic1}. Clearly, the energetically far off extended states play a negligible role for such a deep dot. In (b), the barrier height is reduced to $0.2\hbar v_{\text{F}}q_0$ such that extended states are about as important as bound states.}\label{pic2}
\end{figure}
\begin{figure}[t!]\centering\includegraphics[width = 0.48\textwidth]{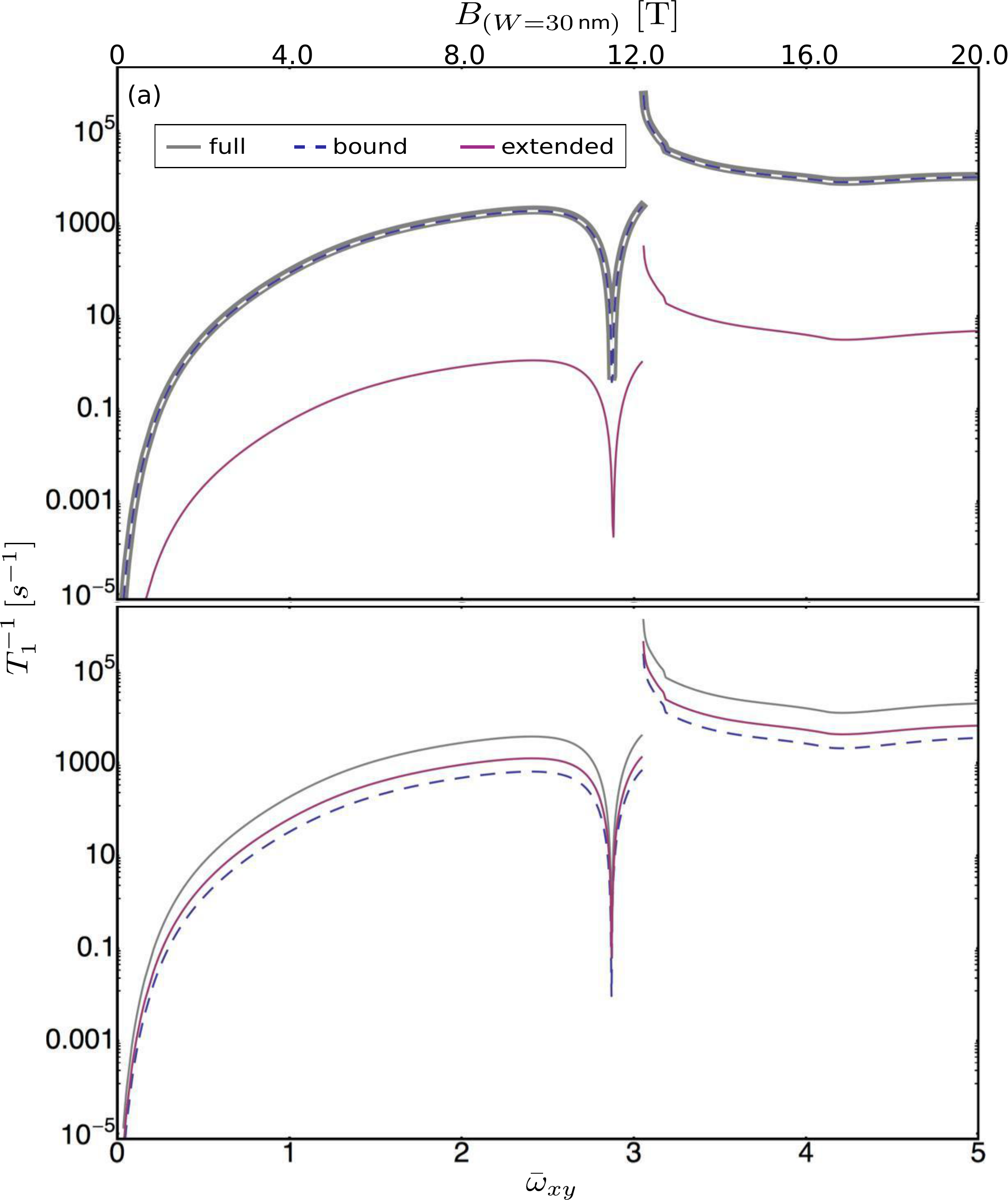}\caption{(Color online) This plot shows the same quantities as Fig.~\ref{pic2}, yet for the aspect ratio $L/W=2$. Again, the influence of extended states depends on the barrier height: $\Delta V=2.0\hbar v_{\text{F}}q_0$ in (a) and $\Delta V=0.9\hbar v_{\text{F}}q_0$ in (b). The extended states dominate in the latter case.}\label{pic3}
\end{figure}

Figure \ref{pic2}(a) shows the full spin relaxation rate for the situation of Fig.~\ref{pic1}, that is, the combined rates of all relaxation channels $\alpha_2$, $\alpha_{4,1}$, $\alpha_{4,2}$, and $\alpha_5$ (gray lines in Fig.~\ref{pic1}) are summed up to the full relaxation rate $T_1^{-1}$ [gray line in Fig.~\ref{pic2}(a)]. The rate with the label ``bound"  (``extended") is obtained in a similar fashion, but only contributions with admixture of bound (extended) states are considered, here. For $\Delta V=1.8\hbar v_{\text{F}}q_0$, the admixture of bound states dominates the admixture of extended states by several orders of magnitude. Yet by lowering $\Delta V$, the influence of extended states can be close to [Fig.~\ref{pic2}(b)] or even surpass the influence of the bound states. Figure \ref{pic3} shows two cases for an aspect ratio of $L/W=2$. In Fig.~\ref{pic3}(a), the barrier height is $\Delta V=2.0\hbar v_{\text{F}}q_0$ and extended states are basically irrelevant compared to the relaxation via bound states. However, Fig.~\ref{pic3}(b) shows that for $\Delta V=0.9\hbar v_{\text{F}}q_0$, the major contribution comes from the extended states.

\begin{figure}[t!]\centering\includegraphics[width = 0.48\textwidth]{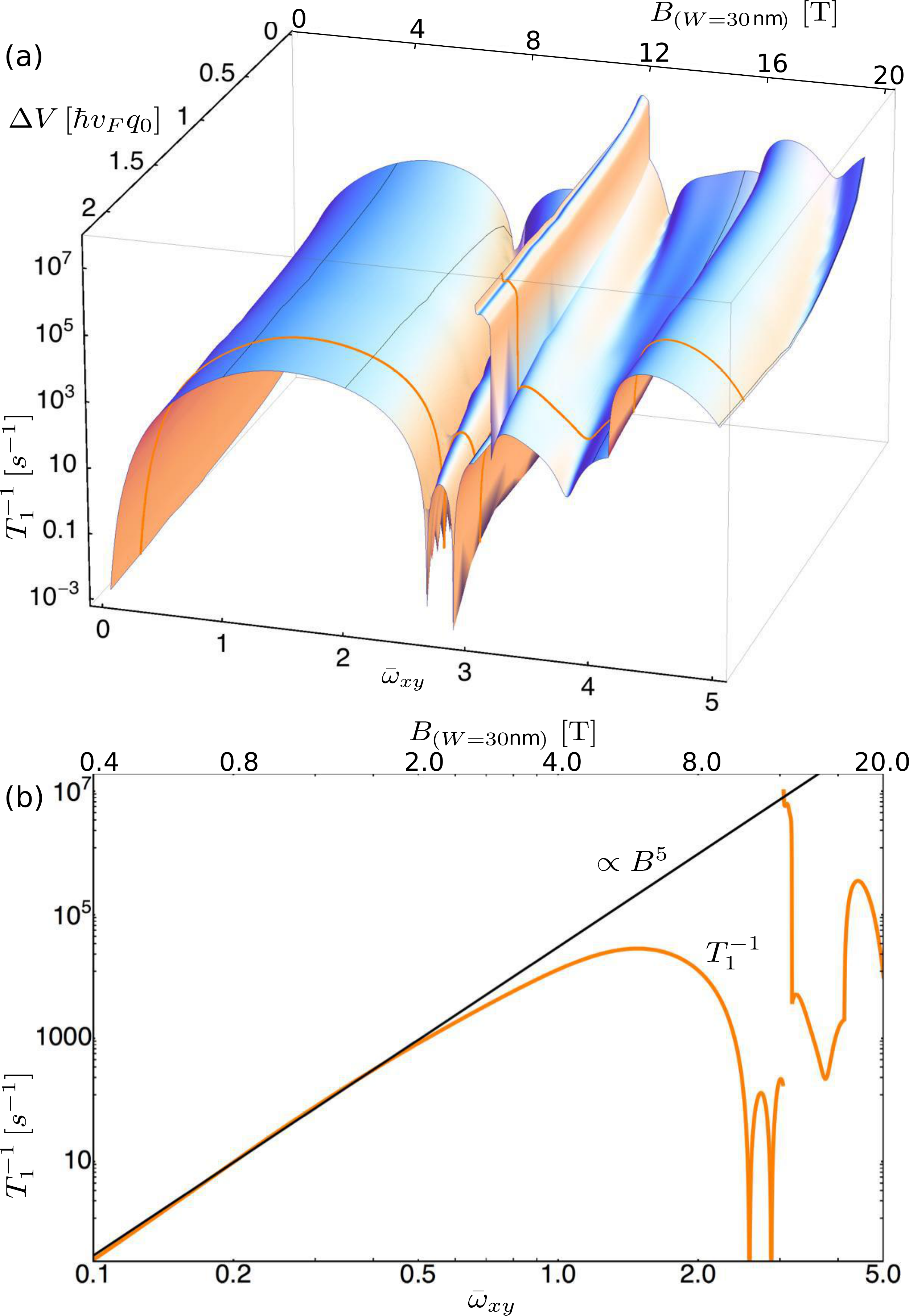}\caption{(Color online) Spin relaxation rate $T_1^{-1}$ for an aGNR with aspect ratio $L/W=5$ and free edges. (a) The rate is shown as a function of barrier height $\Delta V$ and phonon frequency $\bar{\omega}_{xy}$. The orange cut corresponds to the gray line in Fig.~\ref{pic2}(a) and is repeated in (b) with a doubly logarithmic scale that highlights the $B^5$ dependence in the interval $\bar{\omega}_{xy}\in[0.0025,0.5]$.}\label{pic5}
\end{figure}

Figure \ref{pic5}(a) shows $T_1^{-1}$ as a function of parameters $\Delta V$ and $\bar{\omega}_{xy}\propto B$, and for a fixed aspect ratio of $L/W=5$. In contrast to $\bar{\omega}_{xy}$, the barrier height hardly changes the qualitative picture. The orange cut at $\Delta V=1.8\hbar v_{\text{F}}q_0$ is repeated in Fig.~\ref{pic5}(b) in a doubly logarithmic plot that highlights the $B^5$ dependence in the range $\bar{\omega}_{xy}\in[0.0025,0.5]$. In this range, only the branch $\alpha_2$ is available and has a linear dispersion $B\propto\omega\propto q$. The matrix elements $(\mathcal{H}_{\text{EPC}})_{k'k}$ have one power in $B$ due to (i) the gradients $\propto q$ in Eq.~(\ref{epc}), (ii) dipole approximation $\propto q$, and (iii) Van Vleck cancellation ${\propto B}$, each. Because of the prefactor $\propto\omega^{-0.5}$ in Eq.~(\ref{normal}), we find $(\mathcal{H}_{\text{EPC}})_{k'k}\propto B^{2.5}$. As $\alpha_2$ is linear and hence $\rho_{\text{states}}\propto B^0$ for this low-field regime, this explains $T_1^{-1}\propto B^5$. Destructive interference of matrix elements $(\mathcal{H}_{\text{DP}})_{k'k}$ and $(\mathcal{H}_{\text{BLC}})_{k'k}$ can lead to a very small but nonzero relaxation rate.

Figure \ref{pic4}(a) shows a plot similar to Fig.~\ref{pic5}(a), yet for $L/W=2$. The qualitative picture is much different from the aspect ratio $L/W=5$. Figures \ref{pic2} - \ref{pic4}(a) all show discontinuities at $\bar{\omega}_{xy}=3.05$ that stem from the branch $\alpha_4$, for which the density of states has a Van Hove singularity at $\bar{q}_0$ while the coupling $\mathcal{H}_{\text{EPC}}$ remains finite, Fig.~\ref{pic6}(a).

\subsection{Fixed boundary conditions}
\begin{figure}[t!]\centering\includegraphics[width = 0.48\textwidth]{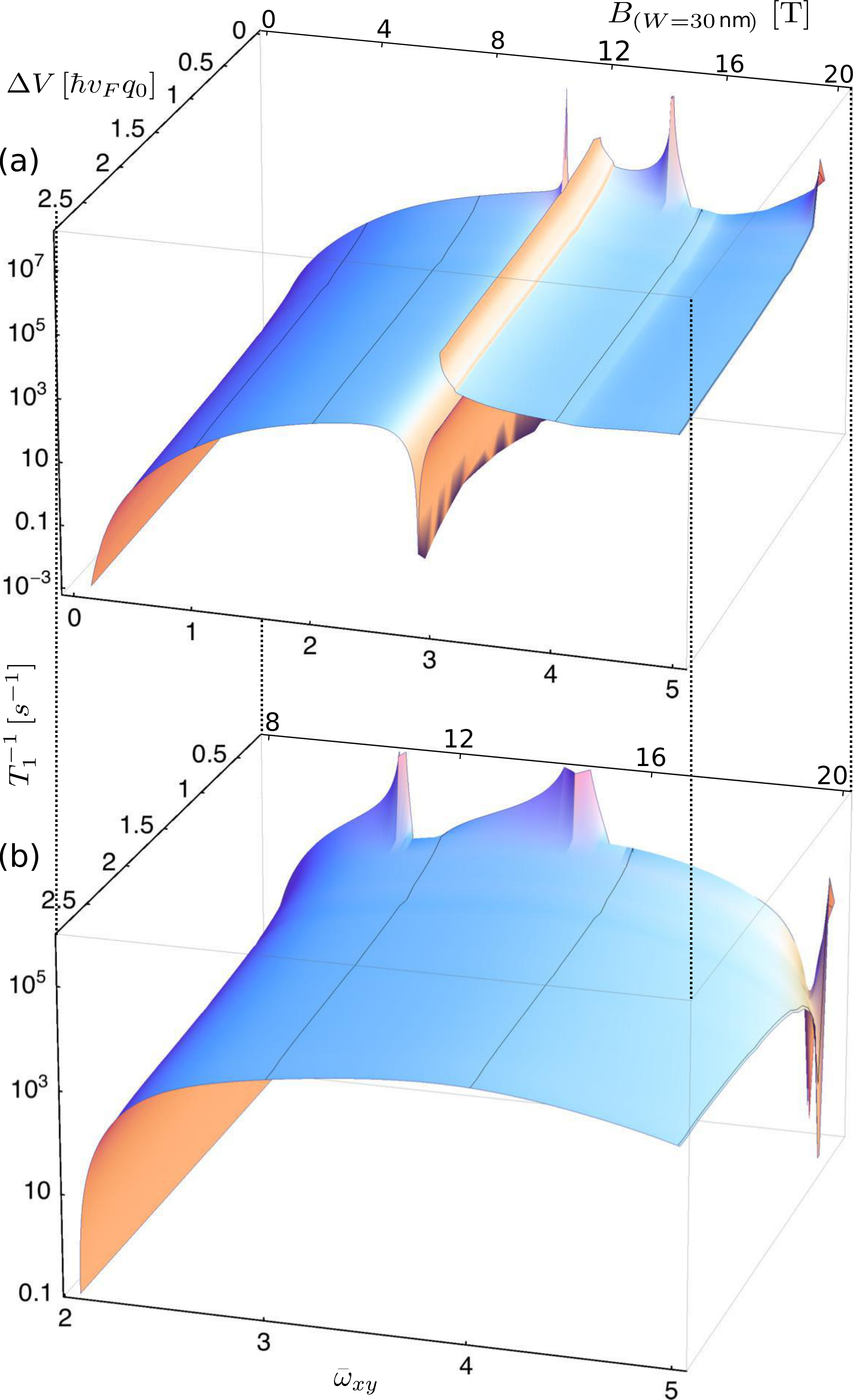}\caption{(Color online) The relaxation rate $T_1^{-1}$ for (a) free and (b) fixed mechanic boundaries. (a) This case is similar to Fig.~\ref{pic5}(a) yet with aspect ratio $L/W=2$. (b) Fixed boundary conditions and $L/W=2$. Due to the gapped phonon spectrum, the rate $T_1^{-1}$ vanishes below $\bar{\omega}_{xy}=2.06$ for our model and with fixed boundaries. Moreover, $T_1^{-1}$ is not discontinuous in $\bar{\omega}_{xy}$ as the branch $\tilde{\alpha}_1$ never becomes flat for finite $\bar{q}$, see Fig.~\ref{pic6}(b).}\label{pic4}
\end{figure}
Most importantly, fixed boundaries result in a gapped phonon spectrum. This means that spin relaxation involving only one phonon cannot occur for magnetic fields that correspond to $\bar{\omega}_{xy}<2.06$. Note, that for a typical width $W=30\,\text{nm}$, $\bar{\omega}_{xy}=2.06$ corresponds to a magnetic field of $8.25\,\text{T}$. However, phonon scattering may still take place below this threshold. In contrast to our claim in Ref.~[\onlinecite{Droth2011}], only the branch $\tilde{\alpha}_1$ contributes to the spin relaxation rate. Gradients, dipole approximation, and Van Vleck cancellation play the same role as for free boundaries, yet due to the gap the frequency $\omega$ is not proportional to some power of $q$ such that there is no power law that connects $T_1^{-1}$ and $B$ as for free boundaries.

Figure \ref{pic4}(b) shows an analog to Fig.~\ref{pic4}(a), yet for fixed boundaries. For aspect ratios larger than in Fig.~\ref{pic4}(b), oscillations occur, which can again be explained with the phonon phase $e^{iqy}$ that rotates according to the phonon dispersion when $\omega$ changes. These oscillations arise only if the dot length is large enough.

\section{Discussion}\label{discuss}
The spin relaxation times we find in our work range from $10^{-7}$ seconds to beyond the second range. For cases where $T_1$ is very long, it can be expected that other mechanisms not considered here will dominate. Our results depend on the aspect ratio $L/W$, the barrier height $\Delta V$, and the Zeeman splitting $g\mu_{\text{B}}B\propto\bar{\omega}_{xy}$ but also on the mechanic boundary conditions that lead to different phonon dispersions. By choosing/adjusting these degrees of freedom properly, $T_1$ can be in the range of seconds. We attribute such long relaxation times to several effects:

(i)
GNRs are quasi one-dimensional systems similar to carbon nanotubes. Both the phonon and the electron density of states are thus limited compared to bulk graphene.\cite{Bulaev2008}

(ii)
Destructive interference between the deformation potential and the bond length change as well as oscillations due to the phonon phase $e^{iqy}$ that rotates according to the phonon dispersion when $\omega$ changes both reduce the relaxation rate $T_1^{-1}$ by several orders of magnitude for specific magnetic fields.

(iii)
In contrast to other graphene QD systems, the electronic states in aGNRs are invariant under time-reversal symmetry, which leads to Van Vleck cancellation.\cite{VanVleck1940,Trauzettel2007,Struck2010} As a result, Eq.~(\ref{admix}) vanishes for ${B=0}$.

(iv)
Deflection coupling to out-of-plane modes vanishes as the evenly distributed weights on $K$ and $K'$ spinor components cancel out. As a result, only the very rigid in-plane modes need to be considered. This rigidity leads to a generally small density of phonon states $\rho_{\text{states}}$.\cite{Trauzettel2007,Rudner2010}

(v)
Phonons do not couple to spin directly so that spin-orbit coupling needs to be included. However, spin-orbit coupling in graphene is weak compared to other systems (e.g. carbon nanotubes).\cite{Trauzettel2007,Bulaev2008,Gmitra2009}

(vi)
The admixture of electronic states in Eq.~(\ref{perturb}) includes bound and extended states. However, only every second bound state contributes; for parity in $y$ direction, $j_k$ and $j_{k'}$ may not be even or odd at the same time, Fig.~\ref{pic7}(b). States that are energetically far apart from the ground state play a small role in the sum which is usually the case for extended states, depending on $\Delta V$. As a consequence, the admixture of these electronic states is suppressed.\cite{Khaetskii2001,Trauzettel2007}

(vii)
Due to parity in the $x$ direction, not all phonon branches contribute to spin relaxation but only those with explicit labels in Fig.~\ref{pic6}, for which Eq.~(\ref{epc}) is even in $x$. This limits the number of relaxation channels.\cite{Droth2011} It is an open question how strong the avoided relaxation channels contribute to $T_1^{-1}$ if this symmetry is broken.

(viii)
We assume phonon vacuum in Eq.~(\ref{T1}). A finite temperature can be included by multiplying the rate $T_1^{-1}$ with the expectation value of the Bose distribution $\langle n_{\alpha,q}(B,T)\rangle$ as explained in Sec.~\ref{eval}.

The carbon isotope $^{12}$C has no nuclear spin and the natural abundance of $^{13}$C, which has spin $1/2$, is only 1\%. Thus, pure dephasing, which comes from coupling to nuclear spins is likely to play a minor role in graphene devices and $T_1\approx T_2/2$ becomes a good measure for overall coherence. Our results show that electronic spin qubits in aGNRs are promising for spintronics applications like the Loss-DiVincenzo quantum computer. With view to recent advances in controlling the edge termination of GNRs it will be interesting to see whether aGNR spintronics can be realized in experiment.
\section{acknowledgements}
We thank the European Science Foundation and the Deutsche Forschungsgemeinschaft (DFG) for support within the EuroGRAPHENE project CONGRAN and the DFG for funding within SFB 767 and FOR 912.

\end{document}